# PERFORMANCE EVALUATION OF QOS IN WIMAX NETWORK

Ahmed Hassan M. Hassan[1,2], Elrasheed Ismail M. ZAYID[3,4]
Mohammed Altayeb Awad[1], Ahmed Salah Mohammed[1], Samreen Tarig Hassan[1]

[1] Mashreq University, Faculty of Engineering, Nourth Khartoum, Sudan
[2] University of Blue Nile, Faculty of Eng., Dept of Electrical Eng., Sudan
[3] Mashreq University, Faculty of Computer Science and IT, Nourth Khartoum, Sudan
[4] University of Elimam Elmahdi, Dept of Computer Eng., Sudan

## ABSTRACT

In this paper, we examine WiMAX – based network and evaluate the performance for quality of service (QoS) using an idea of IEEE 802.16 technology. In our models, the study used a multiprocessor architecture organized by the interconnection network. OPNET Modeler is used to simulate the architecture and to calculate the performance criteria (i.e. throuhput, delay and data dropped) that slightly concerned in network estimation. It is concluded that our models shorten the time quite a bit for obtaining the performance measures of an end-to-end delay as well as throughput can be used as an effective tool for this purpose.

## KEYWORDS

QoS, WiMAX, Performance evaluation

## 1 INTRODUCTION

Parallel computers with multiprocessor systems are opening the door to teraflops computing performance to meet the increasing demand of computational power and can execute the applications faster than a sequential system [1]. Basically parallel computing with multiprocessor systems represents a powerful technology backbone that improves computational productivity by merging a range of standard building nodes and delivering performance across the multi-core platform.

WiMAX is a wireless broadband-broadcasting technology based on wireless metropolitan area networking (WMAN) standard developed by IEEE 802.16 researchers. It provides high data rates, last mile wireless access, point to multipoint communication, large frequency range and guarantees QoS for various applications [2]. The topology of network classified WiMAX into two main sets, IEEE 802.16d-2004 (known as Fixed WiMAX) and IEEE802.16e-2005. It promises to deliver the internet throughout the globe connecting the last mile of communication services [3].

Designing a paralleled-network models are needed to fully exploit hardware capabilities and to achieve the extraordinary energy efficiency target required for science and engineering Applications [2,3].

The study utilized WiMAX which enables the delivery of last mile wireless broadband access as an alternative to wired broadband like cable and digital subscriber line (DSL). Its main advantage is to serve fixed, nomadic and portable wireless broadband connectivity without the need for direct line-of-sight communication path with a base station. Sufficient bandwidth that simultaneously supports hundreds of businesses with T-1 speed connectivity and thousands of residences with DSL speed connectivity. These features candidate WiMAX as a suitable technology for the potential applications as connecting Wi-Fi hotspots with other parts of the internet and providing data and telecommunications services [2,4].

The rest of this article is structured as follows. Section 2 details related to the previous multiprocessor performance evaluation studies. In Section 3 we present overview and architecture of WiMAX. Section 4 provides the simulation framework. Section 5 shows the study findings and results, while the conclusion of this study is presented in Section 6.

## 2 LITERATURE REVIEW

Different system architectures and essays have been proposed to meet the demand for higher performance multiprocessor considering cost and power. Article [4] introduced protocols for broadcasting interconnection network and uses simulation to examine the performance of the protocols over the SOME-Bus multiprocessor architecture. It's successfully applied analytical approach achieving high bandwidth, low latency and large fan-out.

There is a significant study in literature that shows artificial intelligence techniques could be used to predict the performance measures of a multiprocessor architecture [5]. In that study, a broadcast-based multiprocessor architecture called the SOME-Bus employing the DSM programming model was considered. The statistical simulation of the architecture was carried out to generate the dataset. Support vector regression was used to build prediction models for estimating average network response time (i.e. the time interval between the instant when a cache miss causes a message to be enquired in the output channel until the instant when the corresponding data or acknowledge message arrives at the input queue), average channel waiting time (i.e. the time interval between the instant when a packet is enquired in the output channel until the instant when the packet goes under service) and average processor utilization (i.e. average fraction of time that threads are executing). It was concluded that support vector regression model is a promising tool for predicting the performance measures of a DSM multiprocessor. Our basic idea is to collect a small number of performance measures by using a statistical simulation and calculate the performance of the system for a large set of input parameters based on these.

## 3 WiMAX Architectures

Often people access internet wirelessly, either through a mobile handheld devices or from a laptop computer. Today, there are two common types of wireless communication access. In a wireless LAN, wireless users transmit/receive packets to/from an access point that in turn is connected to the main backbone. In wide-area wireless access networks, packets transmitted to a base station over the same infrastructure used for cellular telephony.

Nowadays, WiMAX represents a potential technology hastening communication and dethrone existing standards. WiMAX [Intel WiMAX 2009, WiMAX Forum 2009], also known as IEEE802.16, is a long-distance cousin of the IEEE802.11. WiMAX operates independently of cellular network and promises speeds of tera Mbps or higher over distances of ten kilometers.

Wireless interoperatibility (WiMAX) used for microwave access and can extend the power and range of Wi-Fi and cellular networks. Very often, WiMAX may become the only wireless technology, because Wi-Fi and cellular have not penetrated surfaces that can be reached via WiMAX [4].

Figure1 discusses the general protocol architecture for the IEEE 802.16 standard. As can be seen, a common media access control (MAC) is provided to work on top of different physical layers (PHY). The interface between the different PHYs and the MAC is accommodated as a separate sub-layer, the transmission convergence sub-layer. A convergence sub-layer (CS) is provided on top of the MAC, to accommodate both IP as well as ATM-based network technologies[2].

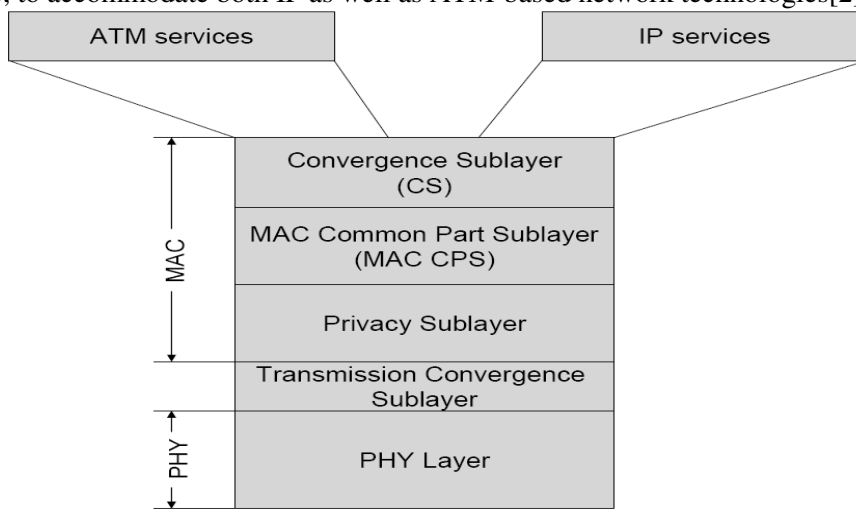

Figure 1: WiMAX Technology Architecture

The following three metrics (throughput, delay and data dropped) are used for evaluating the accuracy of the QoS models**:**

a) Network *throughput* is the average data rate of successful message delivery over a communication channel. This data may be delivered over a physical or logical link, or pass through a certain network node. It is usually measured in bits per second (bit/s or bps) and sometimes in data packets per second or data packets per time slot. This parameter is essential inside the outlook through the device owner and it also measures the amount of user demands that are addressed with the machine. It is an approach to calculating the quantity of service that is provided. The response quantity of confirmed system increases because the system throughput increases. Once the maximum throughput within the techniques accomplished, the response time becomes infinite because the internal queuing delays become arbitrary large.
b) *Delay*: plays an important design and performance characteristic of a computer network or telecommunications network. The delay of a network specifies how long it takes for a

Computer Applications: An International Journal (CAIJ), Vol.2, No.2, May 2015

bit of data travel across the network from one node or endpoint to another. It is typically measured in multiples or fractions of seconds. Delay may differ slightly, depending on the location of the specific pair of communicating nodes. There is a certain minimum level of delay that will be experienced due to the time it takes to transmit a packet serially through a link, this is added a more variable level of delay due to network congestion. IP network delays can range from just a few milliseconds to several hundred milliseconds [6].

c) *Data dropped*, measures the average of data packets that dropped due to their propagation through network model layers (i.e. overflow of buffers) and due to their fails to reach to their receiver entities (failure of all retransmissions until retry limit). It calculated as the total size of higher layer data packets (in bits/sec) dropped by the entire network [7].

## 4 Simulation Framework

In this article, OPNET Module [8] is used to simulate the communication network architecture employing the WiMAX protocol. Figures 2, 3, 4 and 5 show the nodal model of the four different scenarios, respectively.

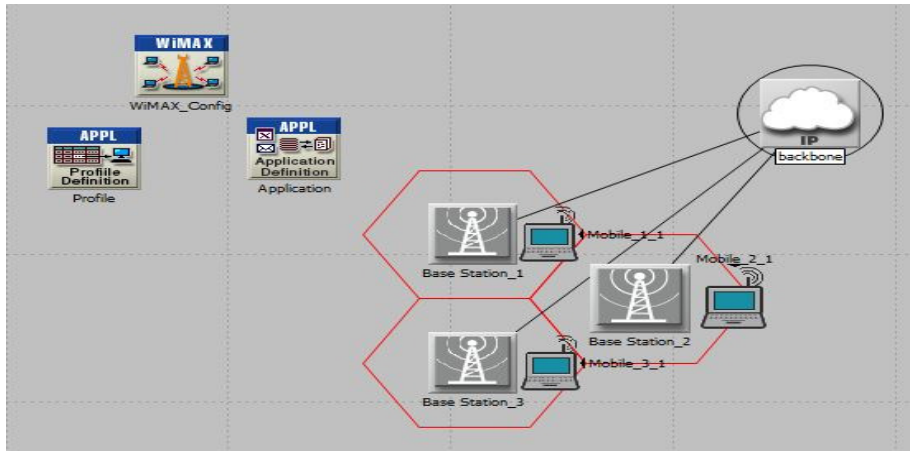

Figure 2: Scenario 1, three base stations with single Mobile.

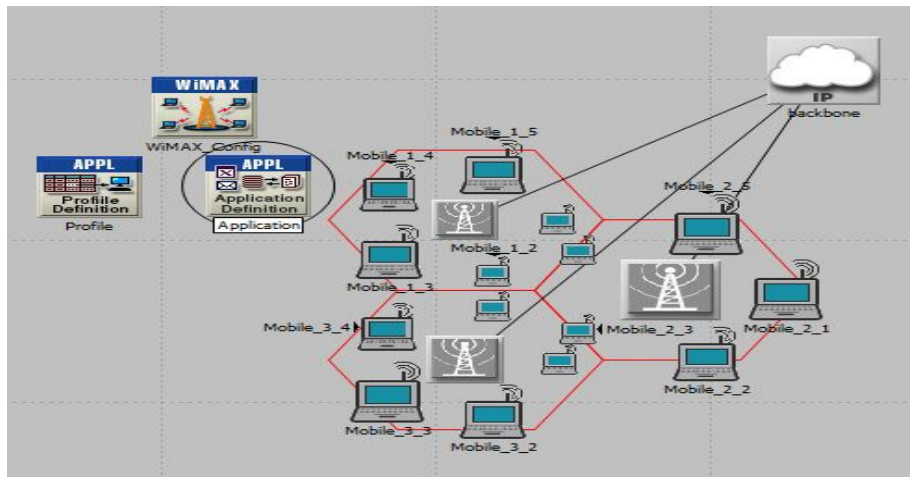

Figure 3: scenario 2, three base stations with three Mobile.

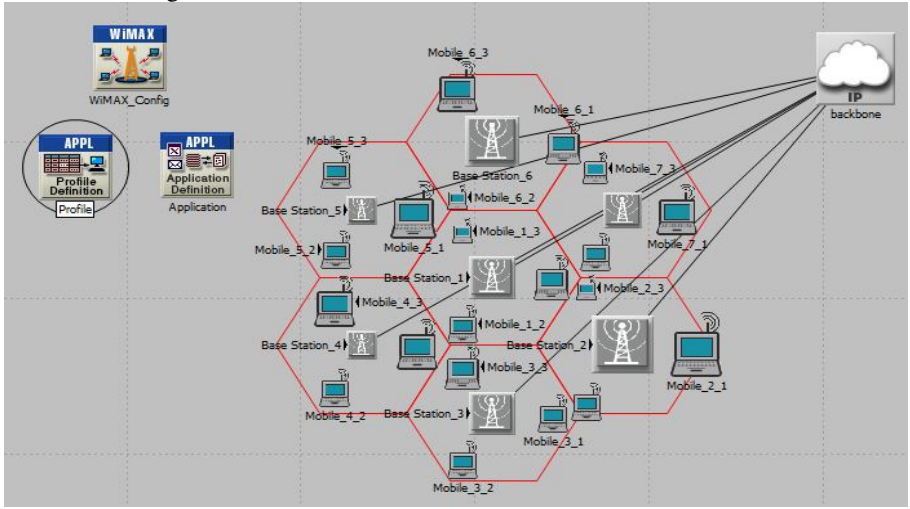

Figure 4: scenario-3, seven cells with three mobile.

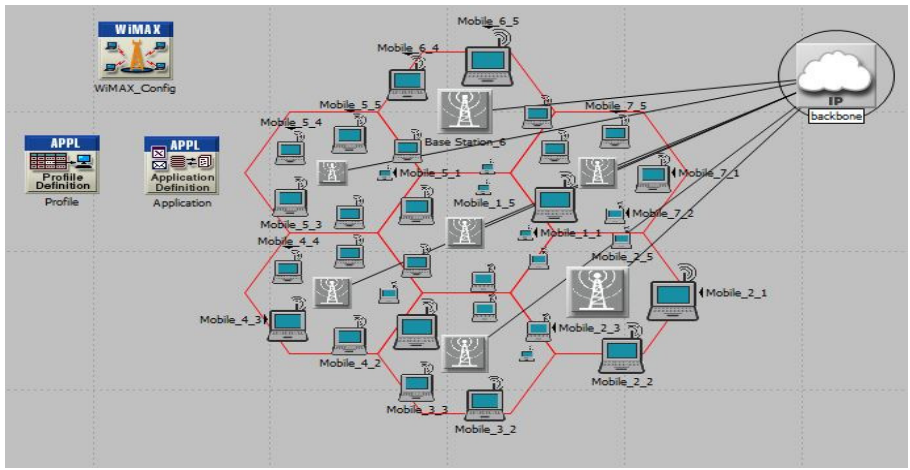

Figure 5: Scenario-4, seven cells with five mobile.

In all Scenarios, each node contains a processor station in which the incoming messages are stored and processed. Furthermore, a channel station in which the outgoing messages are stored before transferring them onto the network.

The important parameters of the simulation are the amount of data transfer successfully performed, the number of the threads executed by each processor (dropped or lost) and the mean delay time for a packet. The output variables are: average throughput, average network response time, average data dropped.

## 5  Results and Discussion:

The simulation has been run for 30 mints and traffic pattern fixed using several quantities to measure and compute the simulation parameters of the QoS. The primary parameters include the averages of: throughput, delay and data drooped as shown in Figures 6, 7 and 8, respectively. The QoS over WiMAX evaluation average rate of decrement/increment rates for all traffic patterns and parameters are summarized in Table 1.

Table 1: The compared accuracy for QoS using our schemes

| No. of Node | Delay | Throughput | Data Dropped |
|---|---|---|---|
| Scenario-1 (3 mobile stations) | No delay | Good throughput | No data dropped |
| Scenario-2 (15 mobile stations) | Slight delay | Better than scenario 1 | Increase in data dropped |
| Sceanrio-3 (21 mobile stations) | Slight delay | Better than scenario 1,2 | Increase in data dropped and worse than scenario 1,2 |
| Scenario-4 (35 mobile stations) | High delay | Much better than all scenarios | Increase in data dropped worse than all scenario |

In the overall models model-4 is a suitable model in applications that require high network traffic performance measures.

From the result shown in Table-1 the following notes can be recommended. In model-1 study utilize three base stations each with a single mobile. The output registered high throughput with zero delay. The bandwidth reservation slightly limits the bandwidth. In model-2 the study uses 3 cells each contains 3 mobiles. Due to free bandwidth, there is an increasing in throughput value with slight delay. The drawback of this scenario that there is no way to enhanced data dropped rate.

In both model-3 and model-4 the numbers of cells increased to seven and the numbers of mobile per cell remains constant in model-3 and increased to five in the later. The bandwidth is established free in both models. In model-3 and model-4, the throughput parameter is acceptable and high, respectively. The delay in model -3 is slightly considered and contrary in model-4. The enhancement of data dropped rate observed in model-4 because increasing the number of base stations meet the traffic needs.

As a result we mentioned that the throughput is rapidly increased compared with the model in Figure 2. Also, for the data dropped it's slightly increased with no significant change observed for delay. In Figure 4 increasing cells number from three to seven boosts up the throughput factor and limits the data dropped, whilst for the delay remains unchanged. In Figure 5 setting the mobile stations by five mobile in each cell we observed that the QoS evaluation parameters: the throughput is quietly increased and both the delay and the data dropped are decreased compared with the other scenarios.

Computer Applications: An International Journal (CAIJ), Vol.2, No.2, May 2015Computer Applications: An International Journal (CAIJ), Vol.2, No.2, May 2015

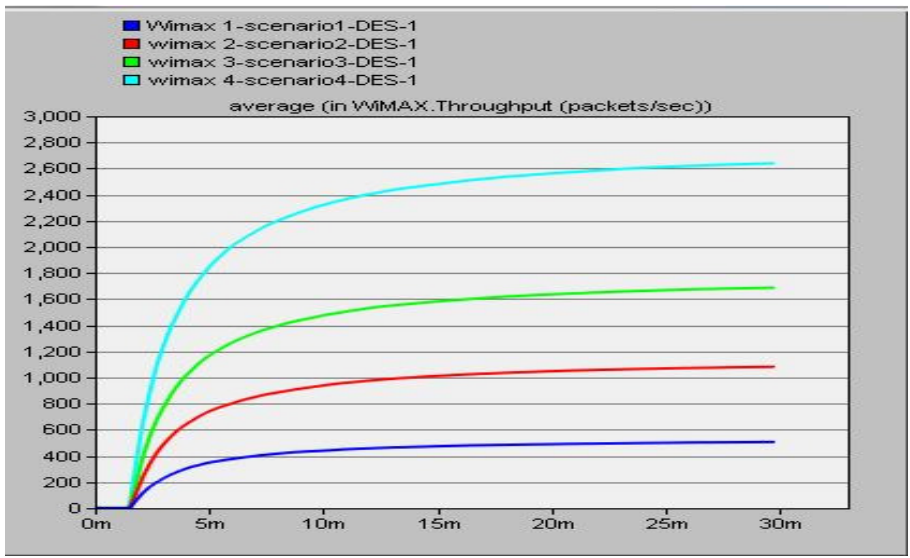

Figure 6: Average throughput computation for the four scenarios over WiMAX

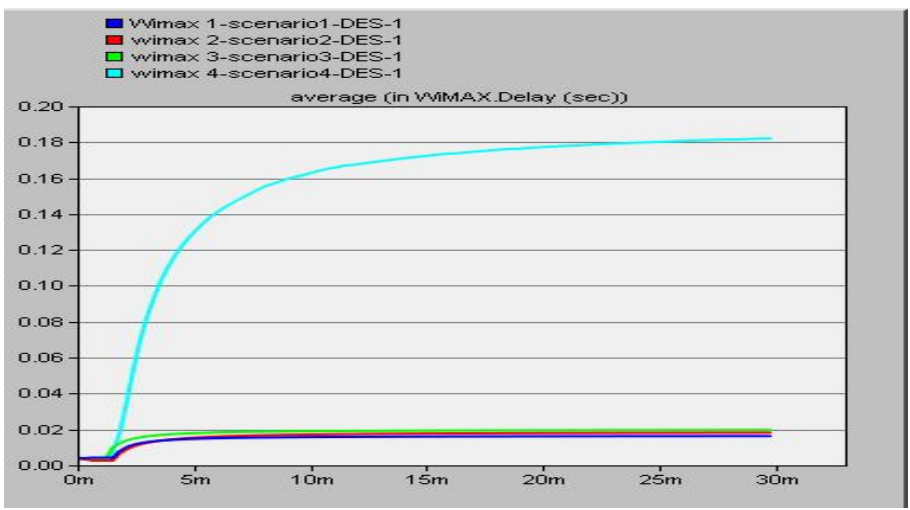

Figure 7: Average Delay computation for the four scenarios over WiMAX

21

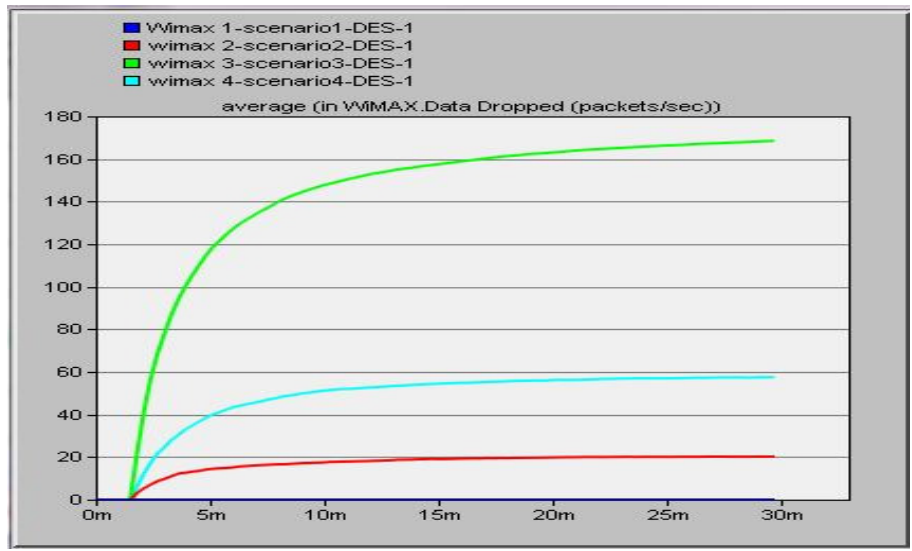

Figure 8: Average Data dropped computation for the four scenarios over WiMAX

## 6 CONCLUSION

This article provides an overview and performance evaluation of QoS in WiMAX network. The study assessed in this paper WiMAX with multiprocessor architecture interconnected by the interconnection network. OPNET Modeler is used to simulate the architecture and to calculate the performance criteria (i.e. *throughput, delay and data dropped)* that slightly concerned in network estimation. The comparison between schemes in terms of performance metrics is provided in Table-1. It is concluded that our models shorten the time quite a bit for obtaining the performance measures of an end-to-end delay as well as throughput can be used as an effective tool for this purpose.

### ACKNOWLEDGEMENTS

The authors would like to thank everyone, just everyone!